\documentclass[12pt]{article}
\usepackage[english]{babel}
\usepackage[dvipsnames,svgnames,x11names]{xcolor}
\usepackage[marginparwidth=2.5cm]{geometry}
 \usepackage[final]{changes}

\usepackage{todonotes}

\usepackage{graphicx}
\usepackage{makeidx}
\usepackage{multirow}
\usepackage{amssymb}
\usepackage{amsfonts}
\usepackage{amsmath}
\usepackage{amsthm}
\usepackage{natbib}

\usepackage{graphicx}
\usepackage{dcolumn}
\usepackage{bm}

\usepackage[utf8]{inputenc}
\usepackage[T1]{fontenc}
\usepackage{mathptmx}
\usepackage{amsfonts}
\usepackage{bbm}
\usepackage[bottom]{footmisc}

\begin{document}

\title
{A new approach towards quantum foundation and some consequences}

\author{Inge S. Helland\\ Department of Mathematics, University of Oslo\\P.O.Box 1053, N-0316 Oslo, Norway\\ ingeh@math.uio.no}

\date{}

\maketitle

\begin{abstract}
A general theory based upon 7 postulates is introduced. The basical notions are theoretical variables that are associated with an observer or with a group of communicating observers. These variables may be accessible or inaccessible. From these postulates, the ordinary formalism of quantum theory \replaced{is}{are} derived. The mathematical derivations are not given in this article, but I refer to the recent articles [9, 10]. Three possible applications of the general theory can be given; 1) The variables may decision variables connected to the decisions of a person or of a group of persons. 2) The variables may be statistical parameters or future data, But most importantly here: 3) The variables are physical variables in some context. This last application gives a completely new foundation of quantum mechanics, a foundation which in my opinion is much more easy to understand than the ordinary formalism. Socalled paradoxes like that of Schr\"{o}dinger's cat can be clarified under the theory. Explanations of the outcomes of David Bohm's version of the EPR experiment and of the Bell experiment are provided. Finally, references to links towards relativity theory and to quantum field theory are given. \added{The concluding remarks point at further possible developments.}

\end{abstract}


\section{Introduction}
\label{intro}

In two books [1, 2] and in a series of articles [3-10] this author has proposed a completely new foundation of quantum theory, and also, in this connection, discussed the interpretation of the theory. This foundation is based upon the general notion of theoretical variables connected to some person or jointly to a communicating group of persons in some given situation. These variables may be accessible or inaccessible, again primitive notions. From a mathematical point of view it is only required that if $\lambda$ is a theoretical variable and $\theta=f(\lambda)$ for some function $f$, then $\theta$ is a theoretical variable. And if $\lambda$ is accessible to the(se) person(s), then $\theta$ is accessible. Some postulates on these variables are stated, and as consequences several quantum-like conclusions are derived.

Depending upon the situation, several different applications of this theory may be given. In a physical context the theoretical variables may be physical variables like time, space, momentum, spin component, energy, mass and charge. A variable is seen to be accessible if it can be measured with arbitrary accuracy. In a statistical context, the variables may be statistical parameters that are seen to be accessible if they can be estimated invariantly with respect to a certain group. Finally, in connection to human decisions, the variables may be decision variables, which are seen to be accessible if the corresponding decision really can be made.

The postulates of the theory will be formulated in the next Section, but I will here mention three of them. 

\replaced{First}{FIrst}, it is assumed that there exists an inaccessible variable $\phi$ such that all the accessible ones can be seen as functions of $\phi$. Group actions are defined on the space upon which $\phi$ varies. In simple physical and statistical situations this can easily be made concrete. 

The first consequence of my postulates are given in Theorem 1 of the next Section, Roughly speaking, this Theorem goes as follows: 

Make some symmetry assumptions; in the discrete case, these seem to be unnecessary. Then assume that there, in some context and relative to a person or to a communicating group of persons, exist two really different maximal accessible theoretical variables. (Using the words of Niels Bohr, these variables may be called complementary.) The term `really different' means that there is no one-to-one function between the variables.

Then the exists a Hilbert space $\mathcal{H}$ connected to the situation, and every accessible varable is connected to a self-adjoint operator in $\mathcal{H}$. It is crucial that the essential assumption behind this is the existence of two complementary variables.

In the discrete case, the eigenvalues of the operator are the possible values of the associated variable. An accessible variable is maximal if and only if all the eigenspaces of the operator are one-dimensional.

The Born rule is, in the physical context, a part of the conclusion. For a detailed derivation of this rule from a few postulates, see [1,10]. In the case where the actual accessible variables are maximal as accessible variables, two postulates seem to be necessary in my derivation.; \added{see Section 2.}

First, the likelihood principle from statistics is assumed (see any textbook in theoretical statistics, or for the origin of this principle, see Berger and Wolpert [11]). To be applicable in a quantum theory situation, a focused version of this principle is derived in [1, 10]. 

The next postulate for the Born rule seems to be important. It is assumed that the relevant observer(s) has/ have ideals, ideals that can be modeled in terms of some abstract being which is seen to be perfectly rational. The rationality is made precise in terms of the Dutch book principle. This postulate may be interpreted in many different directions.

As an interpretation of the whole theory in its physical application, a general epistemic interpretation of quantum mechanics is proposed: Quantum mechanics is not a theory of the world as such, but a theory of an observer's or a communicating group of observers' knowledge about this world. The term `knowledge' is made precise in terms of accessible theoretical variables. In the discrete case, a pure state represents knowledge of the form $\theta=u$ for some maximal accessible variable $\theta$, and, in general, a mixed state represents knowledge in terms of a probability distribution over such a variable.

One aspect of the theory should be mentioned: Since the theory begins with constructing operators connected to variables, these are seen as the fundamental building blocks. Pure states as vectors in $\mathcal{H}$ are only introduced as eigenvectors of physically meaningful operators. This implies a restriction of the Superposition Principle, but note that in the simplest case at least, a qubit, linear combinations of two different state vectors interpreted as spin components, describe new spin components in some directions. Also, it is crucial that certain entangled states vectors also can be given as eigenvectors of some meaningful operator.

The plan of the article is as follows: \replaced{in}{In} Section \ref{sec2} the theory is formulated and made precise. As a first consequence, a discussion of Schr\"{o}dinger's cat paradox is given in Section \ref{sec3}. Then the Bell experiment is briefly discussed in Section \ref{sec4}. Connections to relativity theory and to field theory are mentioned in Section \ref{sec5}. Sections \ref{sec6} and \ref{sec7} give two applications outside quantum foundation, and Section \ref{sec8} gives some concluding remarks.

\section{The basic theory}
\label{sec2}

\subsection{The postulates and the first results}

A completely new approach towards quantum foundations is proposed in Helland [1-10]; see in particular the most recent articles [9] and [10]. The basis can be taken to be an observer who is in some physical situation. In this situation there are theoretical variables, and some of these variables, say $\theta, \lambda, \eta,...$ are related to the observer $C$. Some of these variables are \emph{accessible} to him, which means roughly that it is, in some future, in principle possible to obtain as accurate values as he wishes on the relevant variable. Other variables are \emph{inaccessible}. Two examples of the latter are first the vector (position, momentum) of a particle at some time, and secondly the full spin vector of a spin particle, an imagined vector whose discretized projection in the direction $a$ is the spin component in that direction. 

The above characterization of accessible and inaccessible variables is related to a purely physical implication of the theory. The theory \added{itself} is purely mathematical, can be made precise in different directions, and the terms `accessible' and `inaccessible' are just primitive notions of the theory. Two other ways that the theory can be made precise, are 1) quantum decision theory, where the variables are decision variables; 2) statistical inference theory, where the variables \replaced{either}{eiher} are statistical parameters or future data. From a mathematical point of view, it is only assumed that if $\lambda$ is a theoretical variable and $\theta=f(\lambda)$ for some function $f$, then $\theta$ is a theoretical variable. And if $\lambda$ is accessible, then $\theta$ is accessible.

But the main application here is that these variables are theoretical  variables coupled to some physical situation. If necessary, an observed variable can be modeled as a theoretical variable plus some random error. Following Zwirn [12, 13], every description of reality must be seen from the point of view of some observer. Hence we can assume that the variables also exist relative to $C$. I differ from Zwirn by the fact that I also allow several communicating observers.

\added{(In [1] the theoretical variables were called epistemic variables or e-variable. In some articles I have used the term conceptual variables. I apologize for this confusion.)}

Also, note that observers may communicate. The mathematical model developed in the articles mentioned above is equally valid relative to a group of people that can communicate about the physics and about the various theoretical variables. This gives a new version of the theory, a version where all theoretical variables are defined jointly for such a group. The only difference here is that, for the variables to function during the communication, they must always be possible to define them in words. To be precise, I should also state what the observers should be able to communicate about. In my theory, this is everything that is related to the relevant theoretical variables.

In the two examples mentioned above there are also maximal accessible variables: In the first example this can be either position or momentum, in the second example it can be the spin component $\theta^a$ in some direction $a$. From a mathematical point of view, an accessible variable $\theta$ is called maximal if there is no other accessible variable $\lambda$ such that $\theta = f(\lambda)$ for some non-invertible function $f$. In other words, the term `maximal' will then be seen to be maximal with respect to the partial ordering of variables given by $\alpha\le\beta$ iff $\alpha=f(\beta)$ for some function $f$.

Variables that are maximal, different and not in one-to-one correspondence, are my interpretation of what Niels Bohr called complementary variables.

A basic assumption in my theory is the there exists an inaccessible variable $\phi$ such that all the accessible variables can be seen as functions of $\phi$. In the two mentioned examples, $\phi$ can be taken as the vector (position, momentum), respectively the imagined full spin vector.

\added{Note that my basic article [9] comes in two version. In the arXiv version, some simple category theory is used. This might be the beginning of something important. But the version in \textit{Foundations of Physics} without category theory is equally important. This version has applications for instance in quantum decision theory and in statistics. Note that I only assume that $\phi$ is a variable; its values are not important at all.}

Two different acccessible variables $\theta$ and $\eta$ are defined to be \emph{related} if there is a transformation $k$ in $\phi$-space and a function $f$ such that $\theta = f(\phi)$ and $\eta = f(k\phi)$. 

Two spin components $\theta^a$ and $\theta^b$ are related, and position and momentum are related theoretical variables. In the first case, $\phi$-space can be taken as the plane spanned by the directions $a$ and $b$, and $k$ can be taken as a $180^o$ rotation around the midline between $a$ and $b$. In the last case, $k$ is constructed by a Fourier transform.

As a summary of the above discussion, here are the first 3 postulates of the theory:
\bigskip

\textbf{Postulate 1:} \textit{If $\eta$ is a theoretical variable and $\gamma = f(\eta)$ for some function $f$, then $\gamma$ is also a theoretical variable.}

\bigskip

\textbf{Postulate 2:} \textit{If $\theta$ is accessible to $A$ and $\lambda= f (\theta)$ for some function $f$ , then $\lambda$ is also accessible to $A$.}
\bigskip

\textbf{Postulate 3:} \textit{In the given context there exists an inaccessible variable $\phi$ such that
all the accessible ones can be seen as functions of $\phi$. There is a group $K$ acting upon $\phi$.}
\bigskip

As mentioned, in the two examples above, we may take $\phi =\mathrm{(position,\ momentum)}$, and then let $K$ be the Heisenberg-Weyl  group, and $\phi=$ full spin vector, where $K$ now is taken to be the rotation group. In the last example, say in the spin 1/2 case, one can model the discrete spin component $\theta^a$ in direction $a$ as $f(\phi) = \mathrm{sign}(\mathrm{cos}(a,\phi))$. Giving $\phi$ a reasonable distribution here, results in a correct distribution of each $\theta^a$.

A definition is now needed for the fourth postulate:
\bigskip

\textbf{Definition 1. }
\textit{The accessible variable $\theta$ is called \emph{maximal} if $\theta$ is maximal as an accessible variable under the partial ordering defined by $\alpha\le \beta$ iff $\alpha=f(\beta )$ for some function $f$.}
\bigskip

Note that this partial ordering is consistent with accessibility: If $\beta$ is accessible and $\alpha=f(\beta )$, then $\alpha$ is accessible. Also, $\phi$ from Postulate 3 is an upper bound under this partial ordering. 
\bigskip

\textbf{Postulate 4:} \textit{There exist maximal accessible variables relative to this partial ordering. For every accessible variable $\theta$ there exists a maximal accessible variable $\lambda$ such that $\theta$ is a function of $\lambda$.}
\bigskip

Then, in my opinion, two different maximal accessible variables come very close to what Bohr called complementary variables; see Plotnitsky [14] for a thorough discussion.

It is crucial what is meant by `different' here. If $\theta=f(\eta)$, \deleted{where} $f$ is a bijective function,( \replaced{i.e.}{there is} a one-to-one correspondence between $\theta$ and $\eta$), \replaced{then $\theta$ and $\eta$}{they} contain the same information, and they must be considered `equal' in this sense. $\theta$ and $\eta$ are said to be `different' if they are not `equal' in this meaning. This is consistent with the partial ordering in Definition 1. The word `different' is used in the same meaning in the Theorems below.

Postulate 4 can be motivated by using Zorn’s lemma - if this lemma, which is equivalent to the axiom of choice, is assumed to hold - and Postulate 3, but such a motivation is not necessary if Postulate 4 is accepted. Physical examples of maximal accessible variables
are the position or the momentum of some particle, or the spin component in some
direction. In a more general situation, the maximal accessible variable may be a vector, whose components are simultaneously measurable \added{variables.}

Assuming these postulates, the main result of Helland [3, 9] is as follows:
\bigskip

\textbf{Theorem 1} \textit{Consider a context where there are two different related maximal accessible variables $\theta$ and $\eta$. Assume that both $\theta$ and $\eta$ are real-valued or real vectors, taking at least two values.  Make the following additional assumptions:}
\smallskip

\textit{(i) On one of these \added{two} variables, \added{say} $\theta$, there can be defined a transitive group of actions $G$ with a trivial isotropy group and with a left-invariant measure $\rho$ on the space $\Omega_\theta$, \added{the range space of the function $f$ in $\theta=f(\phi)$.}}

\textit{(ii) The range space $\Omega_\theta$ is either finite and has the same number as $\Omega_\eta$, or more generally, $\Omega_\theta$ and $\Omega_\eta$ have the same category, a notion that can be made precice.}

\textit{Then there exists a Hilbert space $\mathcal{H}$ connected to the situation, and to every (real-valued or vector-valued) accessible variable there can be associated a symmetric operator on $\mathcal{H}$.}
\bigskip

The Hilbert space $\mathcal{H}$ can be taken as $L^2 (\Omega_\theta ,\rho)$, see [42]. The main result is that each accessible variable $\xi$ is associated with an operator $A^\xi$. The proof goes by first constructing $A^\theta$ and $A^\eta$, then operators associated with other accessible variables are found by using the spectral theorem. For \added{a precise definition of the concepts symmetric, self-adjoint and Hermitian and for} conditions under which a symmetric operator is self-adjoint or Hermitian, see Hall [15]; see also [42].
\bigskip

An important special case is when the accessible variables \deleted{that} take a finite number of values, say \replaced{$u_1,u_2,...,u_r$}{$r$}.  For this case it is proved in Helland [9]  that a group $G$ and a transformation $k$such that $\eta(\phi)=\theta(k\phi)$ always can be constructed. The following Corollary then follows: 
\bigskip

\textbf{Corollary 1} \textit{Assume that there exist two different maximal accessible variables $\theta$ and $\eta$, each taking $r$ values, and not in one-to-one correpondence. Then, there exists an $r$-dimensional Hilbert space $\mathcal{H}$ describing the situation, and every accessible variable in this situation will have an associated self-adjoint operator in $\mathcal{H}$.}
\bigskip

Theorem 1 and its Corollary constitute the first step in a new proposed foundation of quantum theory. 

The second step now is to prove the following: If $k$ is the transformation connecting two related maximal accessible variables $\theta$ and $\eta$, and $A^\theta$ and $A^\eta$ are the associated operators, then there is a unitary matrix $W(k)$ such that $A^\eta = W(k)^{-1} A^\theta W(k)$. This, and a more general related result is proved as Theorem 5 in Helland [9].

Given these results, a rich theory follows. The set of eigenvalues of the operator $A^\theta$ is identical to the set of possible values of $\theta$. The variable $\theta$ is maximal if and only if all eigenvalues of the corresponding operator are simple. In general, the eigenspaces of $A^\theta$ are in one-to-one correspondence with questions `What is $\theta$'/ `What will $\theta$ be if we measure it?' together with sharp answers $\theta=u$ \added{for some eigenvalue $u$ of $A^\theta$.}

If $\theta$ is maximal as an accessible variable, the eigenvectors of $A^\theta$ have a similar interpretation. In my opinion, such eigenvectors, where $\theta$ is some meaningful variable, should be tried to be taken as the only possible state vectors. These have straightforward interpretations, and from this version of the theory, also a number of so-called `quantum paradoxes' can be illuminated, see Helland [2, 9]\added{; see also Section 3 below.}

This version of quantum theory implies a restriction of the superposition principle. But note the following: At least in the simplest case, a qubit, where the states can be interpreted as spin components of a particle, a linear combination of two pure states also can be given the same interpretation, i.e. is an eigenvector of some spin operator. Also, the singlet state constructed from two accessible variables, a state that is important both in David Bohm's version of the EPR discussion and in any discussion of the Bell experiment, is an entangled state that is included in my theory. This will be further discussed below.

A final postulate is needed to compute probabilities of independent events. A version of such a postulate is
\bigskip

\textbf{Postulate 5}

\textit{If the probability of an event $E_1$ is computed by a probability amplitude $z_1$ from the Born rule in the Hilbert space $\mathcal{H}_1$, the probability of an event $E_2$ is computed by a probability amplitude $z_2$ from the Born rule in the Hilbert space $\mathcal{H}_2$, and these two events are independent, then the probability of the event $E_1 \cap E_2$ can be computed from the probability amplitude $z_1z_2$, associated with the Hilbert space $\mathcal{H}_1\otimes\mathcal{H}_2$.}
\bigskip

This postulate can be motivated by its relation to classical probability theory: If $P(E_1)=|z_1|^2$ and $P(E_2)=|z_2|^2$, then

\[ P(E_1\cap E_2)=P(E_1)P(E_2)=|z_1|^2 |z_2|^2 = |z_1z_2]^2.\]
\bigskip 

What is lacking in the above theory, is a foundation of Born's formula, \replaced{necessary for the computation of}{the means for calculating} quantum probabilitities. Several versions of Born's formula are proved from two new postulates in Helland [10]. The first postulate is as follows:
\bigskip

\textbf{Postulate 6:} \textit{The likelihood principle from statistical theory holds.}
\bigskip

The likelihood principle is a principle that most statisticians base their inference on, at least when the principle is restricted to a specific context. For basic discussions, see Berger and Wolpert [11], Helland [1] and Schweder and Hjort [43]. It is closely connected to the notion of a statistical model: In an experimental setting, there is a probability model of the data $z$, given some parameter $\theta$. Note that a statistical parameter, in an important application of my theory, can be seen as an accessible theoretical variable. The statistical model can be expressed by a point probability $p$ for discrete data or a probability density $p$ for continuous data: in both cases we have $p=p(z|\theta)$. 

The \emph{likelihood} is defined as $p$, as seen as a function of the parameter $\theta$: 
\begin{equation}
L(\theta ; z)=p(z|\theta).
\label{likelihood}
\end{equation}
\replaced{Then}{and} the likelihood principle runs as follows: \emph{Relative to any experiment, the experimental evidence is always a function of the likelihood.} Here, the term `experimental evidence' is left undefined, and cam be made precise in several directions.

In a \replaced{QM}{quantummechanical} setting, a potential or actual experiment is seen in relation to an observer $C$ or to a communicating group of observers. Concentrate here on the first scenario. In the simplest case we assume that $C$ knows the state $|a;i\rangle$ of a physical system, and that this state can be interpreted as $\theta^a = u_i$ for some maximal accessible variable $\theta^a$. Then assume that $C$ has focused upon a new maximal accessible variable $\theta^b$, and we are interested in the probability distribution of this variable.

The last postulate is connected to the scientific ideals of $C$, ideals that either are given by certain conscious or unconcious principles, or are connected to some concrete persons. These ideal are then \replaced{shaped}{modeled} by some higher being $D$ that $C$ considers to be perfectly rational.

\bigskip

\textbf{Postulate 7:}
\textit{Consider in the context $\tau$ an epistemic setting where the the likelihood principle from statistics is satisfied, and the whole
situation is observed by an experimentalist $C$ whose decisions can be \replaced{shaped or influenced}{modeled to be influenced} by a superior being $D$. Assume that one of $D$'s probabilities for the situation is $q$, and that $D$ can be seen to be perfectly rational in agreement with the Dutch Book Principle.}
\bigskip

The Dutch Book Principle says as follows: \replaced{no}{No} choice of payoffs in a series of bets shall lead to a sure loss for the bettor.

A situation where Postulate 7  holds will be called a \textit{rational epistemic setting}. It will be seen to imply essential aspects of quantum mechanics. As shown in [5], it also gives a foundation for probabilities in quantum decision theory.

In Helland [1,10], a generalized likelihood principle is proved from the ordinary likelihood principle: Given some experiment, or more generally, some context $\tau$ connected to an experiment, any experimental evidence will under the above assumptions be a function of the so-called likelihood effect \replaced{$F=F^a$ or $F^b$, defined by}{; a definition and a discussion is given in [1,10].} 
\added{
$F^a(\bm{u}; z,\tau)=\sum_i p(z|\tau, \theta^a=u_i)|a;i\rangle\langle a;i|.$}

In particular, the probability $q$ is a function of $F$: $q(F|\tau)$.

\subsection{The Born rule}

Using these postulates and a version of Gleason's Theorem due to Busch [16], the following variant of Born's formula is proved in [1,10]:
\bigskip

\textbf{Theorem 2} [Born's formula, simple version] \textit{ Assume a rational epistemic setting \added{and assume two discrete maximal accessible variables $\theta^a$ and $\theta^b$}. In the above situation we have:}
\begin{equation}
P(\theta^b =v_{j} |\theta^a =u_{i})=|\langle a;i|b;j\rangle|^2 .
\label{Born}
\end{equation}

Here, \added{$|a;i\rangle$ is the state given by $\theta^a=u_i$ and} $|b;j\rangle$ is the state given by $\theta^b=v_j$. In this version of the Born formula, I have assumed \emph{perfect measurements:} there is no experimental noise, so that the experiment gives a direct value of the relevant theoretical variable.

An advantage of using the version of Gleason's theorem due to Busch [16] in the derivation of Born's formula, is that this version is valid also in dimension 2. Other derivations using the same point of departure, are Caves et al. [17] and Wright and Weigert [18]. In Wright and Weigert [19] the class of general probabilistic theories which also admit Gleason-type theorems is identified. But for instance, Auffeves and Granger [20] derive the Born formula from other postulates.
\bigskip

The simple Born formula formula can now be generalized to the case where the accessible variables  $\theta^a$ and $\theta^b$ are not necessarily maximal. There is also a variant for a mixed state involving $\theta^a$.

For completeness, define first the mixed state associated with any accessible variable $\theta$. If $\theta$ is not maximal, we need, for the Born formula, the assumption that there exists a maximal accessible variable $\eta$ such that $\theta=f(\eta)$ and such that each distribition of $\eta$, given some $\theta=u$, is uniform. Furthermore some probability distribution of $\theta$ is assumed. Let $\Pi_u$ be the projection of the operator of $\theta$ upon the eigenspace associated with $\theta=u$. Then define in the discrete case the mixed state operator
\begin{equation}
\rho =\sum_j P(\theta=u_j)\Pi_{u_j}=\sum_i\sum_j P(\eta=v_i|\theta=u_j=f(v_i))P(\theta=u_j)|\psi_i\rangle\langle\psi_i |.
\label{Born2}
\end{equation}
Here, $|\psi_i\rangle$ is the state vector associated with the event $\eta=v_i$ for the maximal variable $\eta$\added{, where $v_i$ is an eigenvalue of the corresponding operator $A^\eta$.}

From this definition, we can show from (\ref{Born}), assuming that the maximal $\eta^a$ corresponding to $\theta^a$ also is a function of $\phi$, that in general \added{for a mixed initial state $\rho^a$}
\begin{equation}
P(\theta^b =v|\rho^a)= \mathrm{trace}(\rho^a\Pi_v^b),
\label{Born3}
\end{equation}
with the projection $\Pi_v^b$ projects upon the subspace of $\mathcal{H}$ given by $\theta^b = v$.

This result is not necessarily associated with a microscopic situation. A macroscopic application of Born's formula is given in [1]. \added{See also Section 8.}

The result can also be generalized to continuous theoretical variables by first approximating them by discrete ones. For continuous variables, Born's formula is most easily stated on the form
\begin{equation}
E(\theta^b|\rho^a)=\mathrm{trace}(\rho^a A^{\theta^b}).
\label{Born1}
\end{equation}

Here $E$ denotes the expectation, and $A^{\theta^b}$ is the operator corresponding to $\theta^b$.
We also have
\begin{equation}
P(\theta^b \in B |\rho^a)=\mathrm{trace}(\rho^a \Pi^b(B)),
\label{Born1a}
\end{equation}
where $\Pi^b(B)$ is the projection upon the eigenspace of $A^{\theta^b}$ corresponding to the variable $I(\theta^b \in B)$.

\subsection{The Born rule with data}

Finally, one can generalize to the case where the final measurement is not necessarily perfect. Let us assume future data $z^b$ instead of a perfect theoretical variable $\theta^b$. Strictly speaking, for this case the focused likelihood principle is only valid under the following condition: $p(z^b |\theta^b =u_j)=p(z^b|\theta^b=u_k)$ implies $u_j=u_k$; see [1]. But this is not needed here; we only need the focused likelihood principle for perfect experiments, in order to prove that (\ref{Born3}) is valid. Then we can define an operator corresponding to $z^b$ by
\begin{equation}
A^{z^b}=\sum_j z^b p(z^b |\theta^b =u_j)\Pi_{u_j}^b,
\label{Born4}
\end{equation}
and, from the version (\ref{Born3}) of the Born formula, we obtain
\begin{equation}
E(z^b|\rho^a)=\mathrm{trace}(\rho^a A^{z^b}).
\label{Born5}
\end{equation}

All this not only points at a new foundation of quantum theory, \replaced{but}{and} it also suggests a general epistemic interpretation of the theory: Quantum Theory is not directly a theory about the world, but a theory about an actor's knowledge of the world. Versions of such an interpretation already exist, and they are among the very many suggested interpretations of quantum mechanics. Further discussions of this interpretation are given in [1, 9, 10].

\section{Schr\"{o}dinger's cat}
\label{sec3}

Schr\"{o}dinger's cat is a well known so-called paradox in quantum theory, and there is a large literature on this paradox. It assumes as a thought experiment that a cat is closed into a sealed box, and that there in this box also is some deadly poison that can be released by the decay of some radioactive material. The question that troubled the originator Schr\"{o}dinger and many other physicists was: \deleted{At a time $t$, to an outside observer,} is the cat dead or alive \added{at a given time $t$ for an external observer $O$}? The origin of the difficulty is the Superposition Principle:\added{ for any set of pure states $\{|\psi_i\rangle\}$ and any choice of coefficients $\{c_i\}$ the linear combination $\sum_i c_i |\psi_i\rangle$ is also a state.}. In a \replaced{QM}{quantummechanical} state connected to this observer, \deleted{in} a state which is a superpostion of a `dead' state and an `alive' state, the cat seemingly is both dead and alive.

My theory does not assume a general validity of the Superposition Principle. A pure state vector is included in the theory only if it is an eigenvector of some meaningful physical operator.

It is enlightening to see this in connection to a recent article by Maccone [21]. There, an observator $S$ is introduced which is complementary to the property dead/ alive associated with the cat. In my terminology, there should be a theoretical variable $\theta$ connected to $S$. The crucial question is: Can $\theta$ be measured? According to a long discussion in Skitiniotis et al. [22], this can in principle be done, but it requires a tremendous measurement device. In practical terms I would say that $\theta$ is inaccessible to any given human observer.

To put it simply, a state is always connected to the mind of some observer, and in my theory I allow this observer to answer `I don't know' to certain questions, in this case, the question `Is the cat alive at time $t$?' Once the box is opened, the observer can give a precise answer to the corresponding question.

\added{In principle, the box can be opened at any time. A reviewer has stated the opinion that a `don't know' state is no different than a state of superposition. I disagree. A `don't know' state is referred to a concrete question, involving a physically meaningful observable like $\theta$ above. A superposition is more general.}

\added{Assuming a camera in the box that transmits a signal, and an observer who receives the signal later, one can ask when the cat died. The answer to such a question may under certain circumstances be `don't know', related to this observer or to other observers.}

\added{In general it should be remarked that the explanation of this paradox is simpler in terms of theoretical variables attached to an observer (or to a group of communicating observers) than in terms of a formal quantum theory.}

\section{The Bell experiment}
\label{sec4}

The Bell experiment is a well known experiment whose outcome has caused much discussion among theoretical physicists. In 2022, the Nobel prize in physics was given to three physicists that had performed so-called loophole-free variants of this experiment. The conclusion, which has astonished both scientists an laymen, stands firm. It seems like there either is a violation of the principle of locality, or that the ordinary notion of reality has to be abandoned.

Very briefly, the Bell experiment has to do with two observers Alice and Bob that can not communicate. Midways between the two, a pair of entangled particles are sent out, one towards each observer. Particles with spin 1/2 are assumed. Entanglement means that the spin components of the pair is in a singlet state are given by 
\begin{equation}
|\psi\rangle = \frac{|1+,2-\rangle-|1-,2+\rangle}{\sqrt{2}}.
\label{singlet}
\end{equation}
Here, the spin components of the particles are specified in some direction, say the vertical, 1 and 2 refer to the two particles, and - means downwards and + means upwards. (In the most recent experiments, polarization of phtons is measured, but the discussion is similar.)

Alice is given two directions $a$ and $a'$ to choose from, and measures spin components $A$ or $A'$ in these directions. Bob is given two directions $b$ and $b'$ to choose from, and measures spin components $B$ or $B'$ in these directions. The measured spin components  are coded as -1 and +1.

The experiment is repeated many times, and each time Alice and Bob make their choices. As the measured spin components can be seen as random variables in some sense, expectation of products of pairs presumably exist. The so-called CHSH inequality is
\begin{equation}
E(AB)+E(A'B)+E(AB')-E(A'B') \le 2.
\label{CHSH}
\end{equation}

This inequality may be derived, seemingly using only that the assumption that $A, B, A'$ and $B'$ exist simultaneously as random variables taking the values $\pm 1$, by a very simple argument. But, by suitable choices of $a, b, a'$ and $b'$ it can be shown that (\ref{CHSH}) is violated by quantum mechanics.

Now to the conclusion of very many Bell experiments: \replaced{B}{Again b}y choosing $a, b, a'$ and $b'$ in a suitable way, (\ref{CHSH}) is also violated in practice. Somehow, the simple argument or its assumption must be wrong. In the physical literature, the assumpion behind (\ref{CHSH}) is called local realism.

 In Helland [4, 23], a very specific explanation of the outcome of the Bell experiment is given.
 
 First, it is crucial to show that the state (\ref{singlet}) is allowed by my theory. It should be an eigenstate for some operator which is physically meaningful, which in my terminology means that the operator is connected to a theoretical variable that is accessible to some observer. 
 
 Let \replaced{us as a thought experiment imagine an}{the} observer Charlie \replaced{who is observing both Alice and Bob}{be some person who after the experiments are given all the data from Alice and Bob}, \replaced{and tries to make a mental or mathematical model of}{and then is trying to model} the results of the experiment. During an experiment, he may \added{or may not} be able to observe the whole experiment and record settings and responses. To \replaced{such an imagined actor}{him}, both the spin vectors $\phi_A$ \added{for Alice} and $\phi_B$ for \deleted{Alice and} Bob are inaccessible, but it turns out that the dot-product $\xi =\phi_A \cdot \phi_B$ is accessible to him. In fact, he is forced to be in the state given by $\xi=-3$. This can be seen as follows: The eigenvalues of the operator corresponding to $\xi$ are -3 and +1, and the eigenstate corresponding to $\xi=-3$ is just the singlet state (\ref{singlet}). (Exercise 6.9, page 181 in Susskind and Freedman [24].) Thus, being in this singlet state as an observer of the whole experiment, he is forced to have $\xi = -3$.
 
 What does it mean that the dot product of the two vectors is -3? Expressed by the cartesian components that are $\pm 1$, \added{where $x, y$ and $z$ are the cartesian directions,} we must necessarily have $\phi_A^x \phi_B^x = \phi_A^y \phi_B^y = \phi_A^z \phi_B^z =-1$, that is $\phi_B^x = -\phi_A^x$ etc., which implies $\phi_B^a = -\phi_A^a$ in any direction $a$. This is an important conclusion, and it is true for any \added{such imagined} observer Charlie. This explains the result of David Bohm's version of the EPR experiment, an experiment proposed by Einstein et al [25] \added{and discussed by Bohr [29].} \replaced{The}{and a} result that has caused much confusion, \added{and the discussion has continued until today,}
 
 But go back to the Bell experiment. Here my discussion relies on a general mathematical result of my postulates, a result which in the physical application must be of relevance to the mind of any observer, any person. Recall  the definition of related accessible variables \added{and of maximal accessible variables} given in Section \ref{sec2}.
  \bigskip
 
 \textbf{Theorem 3} \textit{Assume two related maximal accessible variables $\theta$ and $\eta$. Then any other variable $\lambda$ which is related to $\theta$, but unrelated to $\eta$ can not be a maximal accessible variable.}
 \bigskip
 
 The proof is given in [4] and [23]. 
 
 With the physical/ psychological application of my theory described in Section \ref{sec2}, this means that $\lambda$ can not be maximally accessible to the mind(s) of any observer(s) (at the same time; if we let time vary, the observer may have many variables in his mind, also unrelated ones.)
 
 Apply this to the argument around the CHSH inequality \added{ $AB+A'B+AB'-A'B'\le 2$, proved by a simple argument assuming the existence of all the variables $A, B, A'$ and $B'$, all assumed to take the values $\pm 1$.} Consider the observer Charlie, and assume that  the variables $\theta=(A,B)$ and $\eta=(A,B')$ are accessible to him. One can show that, at a fixed time, each of these are maximal \added{to Charlie} as accessible variables: Charlie is only able to observe one run at the time. Also, they are related relative to a $\phi$ containing all of $A, B, A'$ and $B'$. (They have the outcome $A$ in common; we can let the transformation $k$ rotate $b$ onto $b'$ and hence $B$ onto $B'$.) Then look in addition to the variable $\lambda = (A',B)$. This is related to $\theta$, but unrelated to $\eta$. If it were accessible to Charlie, it would again be maximal as an accessible variable, but this is impossible by Theorem 3.
 
 \replaced{Thus, no such observer can have in his or her mind}{This, any observer Charlie can not have} the three variables $\theta, \eta$ and $\lambda$ \deleted{in his mind} at the same time \added{$t$}. \emph{As a consequence of this, he\added{/she} is not able to follow the simple argument leading to the CHSH inequality.} More concretely, if he were statistician, he is not able to put up a valid joint probability model for the variables $A, A', B$ and $B'$. Note that Charlie can be any observer, \added{imagined to be in the above situation}.
 
Thus, the result of the Bell experiment can in my opinion be explained by the fact that we all have a limited mind. This has consequences for this experiment, but as a general observation it also has other consequences.

\added{It shall be noted, though, that the description of Charlie above is a thought experiment. He is thought to be at time $t$ in the special situation that he is about to make a mental or mathematical model explaning what he has seen about the responses from Alice and Bob, and in this situation he is not, at a given time, able to think of all the variables $A, B, A'$ and $B'$ simultaneously. Of course, in any other situation, $\theta$ and $\eta$ above are not maximal to us as accessible variables, and we are then able to think of all the above variables.}

This explanation is not related to non-locality, and the same can be said about my explanation of Bohm's version of the EPR experiment discussed above. Neither can it be directly related to non-reality, but this \added{latter} is a deep issue that can only be discussed in relation to interpertations of quantum theory.

\section{Some consequences for relativity theory and quantum field theory}
\label{sec5}

\added{The core of special and general relativity theory is that physical laws should be the same for every pair of observers that are in relative motion with respect to each other. In my reconstruction of quantum theory, I introduce a theory that can be assumed to be the same for \emph{any} pair of observers, irrespective of their mental state and of the prior knowledge that they have. It is all based on theoretical variables, accessible or inaccessible to the the observer(s) in question. In a relativity setting, these variables may be for instance position, time, momentum or energy. The four-vector constructed from position and time transforms transform according to the Poincare group, and so does the four-vector constructed from momentum and energy. These two four-vectors are complementary. To reach general relativity, Einstein's equivalent principle is crucial: for an isolated observer, gravity and acceleration is equivalent.}

The issues of this Section are discussed in the book [2]. \replaced{As just stated, the}{The} notions of accessible and inaccessible variables are also applicable in special and general relativity theory, and much of the theory generalizes. In particular, it is natural to assume that variables related to the interior of a black hole are inaccessible to any observer. Finally, one may extend the notions to accessible and inaccessible fields. Details are given in the book, some of the discussion is given in [6].

\section{Applications to statistical inference}
\label{sec6}

As stated above, statistical inference is based upon a statistical model specified by a probalility function $p(z|\theta)$ for data $z$ and parameter $\theta$. There is a large literature on statistical inference, but in my opinion, something related to possible structures in the parameterspace is missing. For instance, there may be groups acting on the parameters, and one should try to work towards a systematic theory of model reduction.

My approach involving accessible and inaccessible theoretical variables may have something to contribute here. In Theorem 1 I need a transitive group acting on $\theta$ and in Postulate 3 I assume a transitive group acting upon the large and inaccessible $\phi$. If one can find nontransitive groups acting upon parameters, a reasonable policy for model reduction is \emph{to reduce to a suitable orbit of the group.} At least in the case of partial least squares regression [26] this works well. A systematic application of this principle to quantum mechanics has not been discussed yet.

The case of partial least squares (PLS) regression and its possible relation to the theory of Section 2 here, has been systematically treated in Helland [27]. The total model parameter is called $\phi$, and $\theta$ is the reduced regression parameter corresponding to a PLS model with $m$ steps. The conditions of Theorem 1 is shown to hold in the case where $\eta$ is another reduced parameter, so, in particular a Hilbert space exists with an operator $A^\theta$ corresponding to $\theta$. Optimality of the PLS model compared to other model reductions is \replaced{tried to be treated from this point of view}{discussed}.

\deleted{Another application is the following: Soppose that we have just done a PLS regression based on some data. Let the estimated regression parameter be $\widehat{\theta_1}$. Assume then that we will do a similar investigation later, and call the new estimator $\widehat{\theta_2}$. Then we may look upon these two estimators as maximal accessible variables, use Theorem 1 again, and by the Born formula (\ref{Born5}) in principle find a distribution of $\widehat{\theta_2}$, given a posterior distribution of $\widehat{\theta_1}$, taking into account that the same parametric model is used in the two cases.}

Similar applications \added{of my theory to statistics} should be investigated \added{further.}

\section{A foundation of quantum decision theory}
\label{sec7}

The simplest model of a decision process connected to some person $C$ or to a group of communicating persons is as follows: \replaced{let us assume}{Assume} that the choice is between $r$ possible actions $a_1, a_2, ... ,a_r$. Define a decision variable $\theta$ as being equal to $u$ if action $a_u$ is chosen. As stated in the Introduction, such a decision variable can be taken as a theoretical variable, accessible if the decision really can be done. If there are two complementary decision processes at the same time, the conditions of Corollary 1 are satisfied, and this can be taken as a new theoretical foundation of quantum decision theory, a discipline where most of the literature is based upon empirical investigations, see for instance  Busemeyer and Bruza [28]. This foundation is \deleted{further} discussed in [5].

\section{Concluding remarks}
\label{sec8}

\added{There is a very large literature on quantum mechanics, nearly all of it can be seen as based upon the formalism described in von Neumann's classical book  [30]. As is well known, this book is the result of many years of thinking by eminent physicists, where the points of departure were Heisenberg's work on matrix machanics and Schr\"{o}dinger's work on wave mechanics. This book is a reasonable basis, but it is very formal, and the resulting literature is developed in a similar  formal context.}

\added{The present article is independent of this development. I start completely anew. Such independent approaches have also been done by other researchers, see for instance Hardy [31, 32]. I look upon my own basis as being relatively simple.}

\added{My basic assumption that any description of reality should be connected to an observer (or to a communicating group of observers) is inspired by Herv\'{e} Zwirn's philosophy Convivial Solipsism; see [33] and references there. But otherwise my theory can be seen as a completely independent approach to QM.}

\replaced{I will insist that the}{The} ordinary formal basis for quantum mechanics is very difficult to understand. This has lead to very many interpretations, and for people outside the physical community it seems to be nothing but pure formalism.

By contrast, the foundation discussed in this article is given by 7 postulates, most of them fairly easy to understand, also for \replaced{laymen with some mathematical background}{outsiders}. Still, perhaps, interpretation can be discussed, but a general epistemic interpretation seems to be natural. The basis is just theoretical variables, accessible or inaccessible, and connected to an observer or to a group of communicating observers in some physical context. Since basic quantum theory may be derived from these postulates, everything in the literature which is derived from this formalism can in principle also be deduced from the postulates. An exception is that I restrict the superposition somewhat, as discussed above, and also in [9].

The mathematics behind these derivations is not given here; it is all contained in the recent articles [9] and [10].

Socalled paradoxes like Schr\"{o}dinger's cat can be understood using this theory. The results of the EPR type experiments and the Bell experiments can be explained using the theory.

It is also important that this theory has consequences outside basic quantum theory. A foundation of quantum decision theory may be found. Links to statistical inference theory are beginning to appear. Finally, links to special and general relativity and to quantum field theory can be discussed. A thorough discussion \replaced{is}{will be} given in the \deleted{forthcoming} book [2]; a beginning \added{of this} can be found in [6].

\added{The fact that the whole approach is not microscopic, is extremely important. This means that my theory also can be taken as a basis of the so-called quantum-like models, developed by Andrei Khrennikov and collaborators [34-39], which has as an ambition to cover all of empirical science.}

\added{There are several open problems in my theory. First, I have no explicit discussion showing when operators do not commute. Next, I have a fairly general version of the Born rule, but an even more general version exists, using the notion of \emph{instruments}, see for instance Barndorff-Nielsen et al. [40]. This notion is not yet introduced in my theory. Finally, the Schr\"{o}dinger equation is discussed in [1], but a more thorough discussion could have been made. Its relativistic extensions, the Dirac equation and the Klein-Gordon equation, are not discussed at all.}

\added{One topic that is not discussed in the theory, is the well known rule that a quantum state collapses as a result of a measurement. I do not know how this can be solved, but the following idea may possibly lead to a solution: Look at the basic inaccessible variable $\phi$. Assume that the measurement  instrument is fragile in the sense that $\phi$ changes to another value $\phi'$ during measurement. Perhaps some kind of Bayesian analysis can be made in that connection. A similar idea was recently put forward in [41], a paper that admittedly is incomplete.A completely different solution was proposed in [12].}

I am open to any criticism of this theory, but I also have a hope that further research, using my theory as a basis, can be carried out. 

\section*{References}

[1] Helland, I.S, (2021).
\textit{Epistemic Processes. A Basis for Statistics and Quantum Theory.} 2. edition. Springer, Berlin.
\smallskip

[2] Helland, I.S. and Parthasarathy, H. (2024). \textit{Theoretical Variables, Quantum Theory, Relativistic Quantum Field Theory, and Quantum Gravity.} Manakin Press, New Dehli
\smallskip

[3] Helland, I.S, (2022a). On reconstructing parts of quantum theory from two related maximal conceptual variables.  \textit{International Journal of Theoretical Physics} 61, 69.
\smallskip

[4] Helland, I.S. (2022b). The Bell experiment and the limitation of actors. \textit{Foundations of Physics} 52, 55.
\smallskip

[5] Helland, I.S. (2023a). On the foundation of quantum decision theory. arXiv:2310.12762 [quant-ph]
\smallskip

[6] Helland, I.S. (2023b). Possible connections between relativity theory and a version of quantum theory based upon theoretical variables. arXiv: 2305.15435 [physics.hist-ph]
\smallskip

[7] Helland, I.S. (2023c). Quantum mechanics as a theory that is consistent with the existence of God. \textit{Dialogo Conferences and Journal} 10 (1), 127-134.
\smallskip

[8] Helland, I.S. (2023d). A simple quantum model linked to decisions. \textit{Foundations of Physics} 53, 12.
\smallskip

[9] Helland, I.S, (2024a).
An alternative foundation of quantum theory. \textit{Foundations of Physics} \textbf{54}, 3. \added{arXiv: 2305.06727 [quant-ph].}
\smallskip

[10] Helland, I.S. (2024b). On probabilities in quantum mechanics. APL Quantum 1, 036118.
\smallskip

[11] Berger, J.~O., \& Wolpert, R.~L. (1988). \textit{The likelihood principle.} \deleted{Hayward, CA:} Institute of Mathematical Statistics, \added{Hayward, CA.}
\smallskip

[12] Zwirn, H. (2016). The measurement problem: Decoherence and convivial solipsism. \textit{Foundations of Physics} 46, 635-667.
\smallskip

[13] Zwirn, H. (2023). Everett's interpretation and Convivial Solipsism. \textit{Quantum Rep.} 5 (1), 267-281.
\smallskip

[14] Plotnitsky, A. (2013). \textit{Niels Bohr and Complementarity. An Introduction.} Springer, New York.
\smallskip

[15] Hall, B.C. (2013).  \textit{Quantum Theory for Mathematicians.}  Graduate Texts in Mathematics, \textbf{267}, Springer, Berlin.
\smallskip

[16] Busch, P. (2003). Quantum states and generalized observables: A simple proof of Gleason's Theorem. \textit{ Physical Review Letters, 91}(12), 120403.
\smallskip

[17] Caves, C. ~M., Fuchs, C.~A., \& Schack, R. (2002). Quantum probabilities as Bayesian probabilities. \newblock \textit{Physical Review, A65}, 022305.
\smallskip

[18] Wright, V.J. and Weigert, S. (2019). A Gleason-type theorem for qubits based on mixtures of projective measurements. \textit{ journal of Physics A: Mathematical and Theoretical} 52 (5), 055301.
\smallskip

[19] Wright, V.J. and Weigert, S. (2021). General probabilistic theories with a Gleason-type theorem. \textit{Quantum} 5, 588.

[20] Auffeves, A. and Granger, P. (2019). Deriving Born's rule from an inference to the best explanation. arXiv:1910.13738 [quant-ph].
\smallskip

[21] Maccone, L. (2024). Schr\"{o}dinger cats and quantum complementarity. \textit{Foundations of Physics} 54, 17.
\smallskip

[22] Skotiniosis, M., D\"{u}r, W. and Sekatski, P. (2017). Macroscopic superpositions require tremendous measurement devices. \textit{Quantum} 1, 34.
\smallskip

[23] Helland, I.S. (2023e). On the Bell experiment and quantum foundation. arXiv: 2305.05299 [quant-ph]. \textit{J. Mod. Appl. Phys.} 6 (2), 1-5.
\smallskip

[24] Susskind, L. and Friedman, A. (2014) \textit{Quantum Mechanics. The Theoretical Minimum.} Basic Books, New York.
\smallskip

[25] Einstein, A., Podolsky, B. and Rosen, N. (1935). Can quantum-mechanical description of physical reality be considered complete? \textit{Physical Review} 47, 777-780.
\smallskip

[26] Helland, I.S., S\ae b\o , S. and Tjelmeland, H. (2012). Near optimal prediction from relevant components. \textit{Scand. J. Statist.} 39, 695-713.
\smallskip

[27] Helland, I.S. (2024c). Towards optimal linear predictions. arXiv: 2412.19186 [math.ST].
\smallskip

[28] Busemeyer, J.R. and Bruza, P.D. (2012). \textit{Quantum Models of Cognition and Decision.} Cambridge University Press, Cambridge.
\smallskip

[29] Bohr, N. (1935) Can quantum mechanical description of physical reality be considered complete? \textit{Physical Review}, 45, 696-702.
\smallskip

[30] von Neumann, J. (1932). \textit{Mathematische Grundlagen der Quantenmechanik}. Springer, Berlin. Translated: \textit{Mathematical Foundation of Quantum Mechanics}. Princeton University Press, Princeton.
\smallskip

[31] Hardy, L. (2001) Quantum theory from five reasonable axioms. arXiv: 01010112v4 [quant-ph].
\smallskip

[32] Hardy, L. (2013) Reconstructing quantum theory. arXiv: 1303.1538 [quant-ph].
\smallskip

[33] Zwirn, H. (2023) Everett's interpretation and Convivial Solipsism. \textit{Quantum Rep}. 5 (1), 267-281.
\smallskip

[34] Khrennikov, A. (2010). \textit{Ubiquitous Quantum Structure. From Psychology to Finance.} Springer, Berlin.
\smallskip

[35] Haven, E. and Khrennikov, A. (2013). \textit{Quantum Social Science.} Cambridge University Press, Cambridge.
\smallskip

[36] Khrennikov, A. (2023). \textit{Open Quantum Systems in Biology, Cognitive and Social Sciences.} Springer Nature Switzerland.
\smallskip

[37] Haven, E. and Khrennikov, A. (2016). Statistical and subjective interpretations of probability in quantum-like models in cognition and decision-making. \textit{J. Mathematical Psychology} 74, 82-91.
\smallskip

[38]Veloz, T., Khrennikov, A., Toni, B. and Castillo, R.D. [ed.] (2023). \textit{Trends and Challenges} \textit{in Cognitive Modeling.} Springer Nature, Switzerland.
\smallskip

[39] Ozawa, M. and Khrennikov, A. (2021). Modeling combination of question order effect, response replicability, and QQ-equality with quantum instruments. \textit{J. Mathematical Psychology} 100, 102491.
\smallskip

[40] Barndorff-Nielsen, O.E., Gill, R.D. and Jupp, P.E. (2003). On quantum statistical inference. \textit{Journal of the Statistical Society B} 65 (4), 1-31.
\smallskip

[41] Navarrette, Y. and Davis, S. (2024). Fragile systems: A hidden-variable Bayesian framework leading to quantum theory. arXiv: 1609.019972 [quant-ph].
\smallskip

[42] Helland, I.S. (2025). Some mathematical issues regarding a new approach towards quantum foundation. arXiv: 2411.13113 [quant-ph].
\smallskip

[43] Schweder, T. and Hjort, N.L. (2016). \textit{Confidence, Likelihood, Probability. Statistical Inference with Confidence Distributions.} Cambridge University Press, Cambridge.

\end{document}